**Bridging the Gap: Understanding Rural Commuting Patterns and Transportation Choices**

**Demayla Jenkins**
Undergraduate Student, Construction Engineering Technology
School of Architecture and Engineering Technology
Florida Agricultural and Mechanical University, Tallahassee, Florida 32307
Email: demayla1.jenkins@famu.edu

**Janeroza Matyenyi**
Graduate Research Assistant, School of Engineering
University of North Florida, Jacksonville, FL 32224
Email: janerose.matyenyi@unf.edu

**Thobias Sando, Ph.D., PE, PTOE**
Professor, School of Engineering
University of North Florida, Jacksonville, FL 32224
Email: t.sando@unf.edu

**Doreen Kobelo Regalado, Ph.D.**
Director, Division of Engineering Technology
School of Architecture and Engineering Technology
Florida Agricultural and Mechanical Engineering, Tallahassee, FL 32307
Email: doreen.kobelo@famu.edu

**Mohamed Khalafalla, Ph.D., MBA, PMP**
Assistant Professor, Construction Engineering Technology
School of Architecture and Engineering Technology
Florida Agricultural and Mechanical University, Tallahassee, Florida 32307
Email: Mohamed.Ahmed@famu.edu

Word Count: 4,554 words + 3 table (250 words per table) = 5,304 words

*Submitted August 1st, 2024*

*Presented January 2025*

**ABSTRACT**
Transportation provides access to employment opportunities and essential services such as healthcare services; while urban areas have various transportation options, the situation differs in rural areas. Rural residents often have longer commute distances, limited access to public transit, and extended waiting times for public transportation if they exist, which can significantly impact their access to vital services and job opportunities. This study used data from the 2017 NHTS survey to examine the commuting patterns in rural areas by utilizing multinomial logistic regression to determine how various factors impact the choice of mode of transport in rural areas. Findings from this study revealed a higher dependency, 92.1%, on using personal vehicles when making trips in rural areas. Multinomial logistic regression results showed that socio-demographics, household, and trip characteristics affect the mode of transport used in a trip. Older adults, females, and individuals with higher education levels than high school graduates are less likely to use public transit when making trips. For household characteristics, the availability of vehicles in a household and households with higher income levels have lower probabilities of making a trip using public transit. Longer trip distances reduce the likelihood of a trip using active commuting modes such as walking and biking. These findings provide insights into understanding the transportation behaviors in rural areas and provide knowledge to be used in the planning and developing transportation projects that promote equitable and accessible transportation in rural areas.

**Keywords:** Rural transportation, commuting patterns, socio-economic impact on mode choice

**INTRODUCTION**
The significance of having various transport options cannot be overstated; diverse transportation enhances access to education, employment, and healthcare. It connects multiple areas, reduces traffic congestion by promoting public transit, decreases environmental impact, and improves public health by promoting active transportation such as walking and cycling. While urban areas have various transportation options, including public transit, walking, cycling, and ridesharing services, rural areas face significant challenges (*1*). Limited access to transportation services in rural areas forces residents to rely mostly on personal vehicles, which leads to higher transportation costs and reduced accessibility for non-drivers, such as the elderly, low-income individuals, and people with disabilities (*2, 3*). Studies show that rural residents commute longer distances and take more time compared to their urban counterparts.

Analyzing travel patterns in rural communities highlights the unique transportation challenges these areas face. Understanding travel behaviors enables the development of targeted solutions to improve mobility and access to essential services in rural areas. This study examines the commute patterns of rural residents and explores how various factors, including socio-demographic, trip, and household characteristics, influence their mode of transportation. Additionally, it compares commuting characteristics across different census regions and analyzes changes in travel behavior over the years using data from the National Household Travel Survey (NHTS) released in 2017 and 2022.

**LITERATURE REVIEW**
Definitions of rural and urban areas vary across several federal agencies. The National Household Travel Survey (NHTS) utilizes the U.S. Census designation to differentiate between urban and rural residences. This classification defines rural and urban areas according to the United States Census Bureau, which defines urban and rural areas based on population thresholds, land use, density, and housing (*4*). Rural areas are identified as territories, housing, and populations that are not urban or urban clusters. In contrast, an urban area is a densely developed territory comprising residential, commercial, and other non-residential urban land uses. An urban area comprises a densely settled core of census blocks with a minimum housing unit density of 2000, a population density of at least 1000 per square mile, and a population of at least 5000 people (*5*).

The Census Bureau groups the states and counties into various regions and divisions representing different sections of the United States. The current configuration of Census regions was established in 1910, following modifications to the initial regions introduced in 1850. The Census Bureau classifies states and counties into four regions: Northeast, West, Midwest, and South. These regions are further classified into nine divisions: Pacific, Mountain, West South Central, East South Central, South Atlantic, West North Central, East North Central, Middle Atlantic, and New England (*6*).

Several studies have compared the characteristics of commuting in rural and urban areas. In rural regions, daily trips are very few, and these trips often cover longer distances than those undertaken by their urban counterparts. Transportation in rural communities has a higher dependency on private vehicles for daily commuting, contributing to increased average travel distances (*7*). As a result, rural residents bear a more significant out-of-pocket burden for travel expenses, primarily associated with automobile-related costs such as vehicle purchases and gasoline (*3*). Additionally, the limited availability of public transit services in 18 percent of rural U.S. counties further exacerbates the challenges faced by non-urban travelers, who often have limited commuting options (*8*).

Several studies have highlighted the challenges of providing effective public transit in rural settings. Low population densities, dispersed development patterns, and longer travel distances make it difficult to achieve the ridership levels and operational efficiency necessary to sustain robust public transportation systems. Additionally, rural transit agencies often face significant funding constraints, limiting their ability to expand service coverage and increase service frequency. Despite these challenges, some rural communities have found innovative ways to improve public transportation access. Flexible, on-demand micro-transit services, such as dial-a-ride or deviated fixed-route systems, have emerged as a more responsive and cost-effective alternative to traditional fixed-route bus services (*8–10*).

Table 1 summarizes past studies that have used the NHTS data and analyzed various travel trends and behavior in the United States. It summarizes the study objectives, data source, variables used in the study, the methodology, and the findings.

**Table 1. Studies that have used the NHTS data**

| Study | Objective | Data | Methodology | Findings |
|---|---|---|---|---|
| (*11*) | Factors that determine the choice of commuting mode in California (Los Angeles, San Diego, San Francisco, San Jose, and Sacramento) | 2017 NHTS California and spatial data | Analysis using binary logistic regression, comparison among the study area cities | Trip distances encourage car use in all cities, the number of vehicles present in a household also increases usage of personal vehicles, distance to public transit stops decreases the preference for using public transit, and an increase in household size encourages the use of public transit in some cities and negatively affect the use of public transit in other cities. |
| (*12*) | Analyze how travel patterns differ across age groups in America | 2001 NHTS data | Descriptive Statistics | Older Americans rely on personal vehicles for travel; they take fewer trips, travel shorter distances, and have shorter travel times. |
| (*13*) | Identify the differences in work-from-home behavior among rural and non-rural U.S workers | 2017 NHTS data | Binary and multinomial logistic regression | Urban and rural workers differ in terms of working-from-home access. Rural people worked more from home, while urban workers were far more likely to commute by choice. |
| (*7*) | Comparing the travel behavior in rural and urban areas in the U.S. | 2001 NHTS data | Descriptive Statistics | Mobility levels in rural residents are higher than in urban residents. Rural residents make fewer daily trips, trips in rural areas cover long distances, and people in rural areas often use private vehicles. |
| (*14*) | Estimating bicycling prevalence in urban and rural areas | 2017 NHTS data | Logistic regression model | Urban areas have a higher prevalence of bicycling for commuting than rural areas. |
| (*15*) | Effect of built-in Environment on active commuting in rural and urban areas | 2010 Decennial census data and 2011-2017 American Community Survey | Regression models (multivariate logit regression) | Rural tracts had a higher rate of walking to work than public transit or biking; socio-demographic factors explained a large variance in an active commuting mode in rural and urban areas, and environmental variables and active commuting association differed in rural and urban areas. |

**METHODOLOGY**
**Data**
The availability of the 2017 and 2022 NHTS data allows a detailed analysis of this study. The National Household Travel Survey is conducted by the Federal Highway Administration every 5 to 7 years, and most recently, it was conducted in 2022. The NHTS data is the source of data on travel behavior and trends at the national level. The survey collects data on demographic characteristics such as age, gender, and race, data on travel-related characteristics including vehicle types, daily trip making, and distance traveled within the trips, and data on socio-economic factors such as the income of households. The 2017 survey was conducted between April 2016 and April 2017. It includes information on 923,572 trips made by 264,234 individuals aged five or older in 129,696 US households (*16*).

The data consists of four files: the household file, which contains data collected once for the household; the person file, which includes data collected once for each interviewed household member; the vehicle file, which has data related to household vehicles; and the trip file, which has data items collected for each trip made. This study used the trip file, which contains data on all trips made on the assigned date by persons aged five and older, to analyze the travel behavior of rural residents.

Table 2 presents the descriptive statistics for data from 2017 and 2022, revealing a 2.1% increase in personal car usage in 2022 compared to 2017. Walking trips and trips made by other modes decreased from 5.9% to 2.6% and from 1.2% to 0.3%, respectively. Interestingly, trips made using public transit increased from 0.4% to 2.5%. These changes could be attributed to the rise of the pandemic, which led to a preference for modes of transport that offer personal isolation, such as personal vehicles, and a reduction in modes, such as ridesharing and walking. However, the increase in public transit use is counterintuitive.

The data shows a slight change in trip day distribution from 2017 to 2022; weekend trips increased from 22.0% to 25.0%, and weekday trips decreased from 78.0% to 75.0%. This change in trip days coincides with more significant shifts in trip purposes. Work-related trips increased from 13.3% to 19.7%, aligning with the still-high percentage of weekday trips. Recreational trips increased from 10.5% to 23.4%, while shopping and non-home-based trips declined from 20.9% to 11.8% and from 37.0% to 20.9%, respectively. These changes suggest a structured travel pattern in rural areas, with more trips taken on weekdays and work and recreational activities being more prominent reasons for travel.

The distribution of trips by age shows an increase for younger adults, with a 3% rise for the 11-30 age group and a 5.2% increase for the 31-50 age group. In contrast, older adults (51 and above) experienced a decrease in trips. Specifically, trips by adults aged 51-70 decreased from 49.3% to 41.4%, and trips by adults aged 71 and above decreased from 17.2% to 16.8%. This indicates a decline in trips made by older adults in rural areas.

**Multinomial Logistic Regression**
The study employed a Multinomial Logistic Regression (MNL) model to explain the relationship between mode of transport choice and residents' socio-demographic, economic, and trip characteristics in rural areas. The MNL model is a statistical tool that examines a categorical outcome variable with more than two categories about a set of independent variables (*17, 18*). It is typically used to understand how individuals or organizations make decisions (*19*). The MNL model assumes that variations in decision-maker characteristics lead to variations in choice outcomes(*20*). Suppose the outcome variable Y has K number of observed categories. Equation (1) shows the probability of each choosing each category from the response variable.

$$Prob(Y = i) = \frac{exp\,(\lambda_i)}{\sum_{h=1}^{K} exp\,(\lambda_{hi})} \quad (1)$$

where:
$\lambda_i = \beta_0 + \sum_{h=1}^{H} \beta_{ih} X_{ih}$
$\beta_0$: the constant term.
$\beta_{ih}$: the vector of estimable coefficients.
$X_{ih}$: the vector of explanatory variables.

**Table 2. Descriptive statistics**

| Variable | Category | Percentage (%) 2017 | Percentage (%) 2022 |
|---|---|---|---|
| Age | 11-30 | 11.1 | 14.1 |
| | 31-50 | 22.5 | 27.7 |
| | 51-70 | 49.3 | 41.4 |
| | 71 and above | 17.2 | 16.8 |
| Gender | Male | 47.7 | 50.3 |
| | Female | 52.3 | 49.7 |
| Education level | Highschool graduate or less | 29.5 | 24.1 |
| | College degree | 31.6 | 34.2 |
| | Bachelor's degree | 38.8 | 41.7 |
| Race | White | 89.7 | 91.8 |
| | Black | 4.0 | 2.6 |
| | Hispanic | 2.9 | 2.8 |
| | Other | 3.4 | 2.8 |
| Household Income | less than 25,000 | 14.0 | 7.7 |
| | 25000-49999 | 22.1 | 14.5 |
| | 50,000-74999 | 20.0 | 18.3 |
| | 75,000-99999 | 15.5 | 21.1 |
| | 100,000 and more | 28.5 | 38.5 |
| Trip purpose | Work | 13.3 | 19.7 |
| | Non-home based | 37.0 | 20.9 |
| | Shopping | 20.9 | 11.8 |
| | Recreation | 10.5 | 23.4 |
| | Other | 18.2 | 24.2 |
| Trip Day | Weekend | 22.0 | 25.0 |
| | Weekday | 78.0 | 75.0 |
| Census Region | Northeast | 21.7 | 15.7 |
| | Midwest | 18.0 | 33.7 |
| | South | 45.5 | 36.3 |
| | West | 14.8 | 14.3 |
| Mode of Transport | Personal Car | 92.1 | 94.2 |
| | Transit | 0.4 | 2.5 |
| | Walk | 5.9 | 2.6 |
| | Bike | 0.3 | 0.3 |
| | Other | 1.2 | 0.4 |

Data preprocessing steps were carried out prior to model fitting. This process included handling missing values, coding categorical variables correctly, and assessing multicollinearity among the independent variables. Numerical variables such as household size, number of household vehicles, and trip miles were retained to enhance the prediction accuracy of the model. The category "other" for modes of transport included taxis, uber/Lyft, rental cars, golf carts, and any modes described as "something else" in the data.

Trips involving airplanes and water ferries were excluded from the analysis. Highly correlated variables were identified, and one of each pair of correlated variables was excluded to prevent multicollinearity issues. For example, since trip duration was highly correlated with trip distance, it was removed from the analysis.

The model was fitted using R software with the "nnet" package. The maximum likelihood estimation method, which identifies the regression coefficients that maximize the likelihood of the observed data, was employed to estimate model parameters. The model's goodness-of-fit was evaluated using the deviance statistic, which compares the fitted model to a saturated model. Results were interpreted by examining each variable's odds ratio and p-values. Coefficients provided insights into the direction and magnitude of each independent variable's impact on the likelihood of selecting a particular category in the response variable.

**RESULTS AND DISCUSSION**
Results from the multinomial logistic regression are presented in Table 3. Modes of transport were categorized as public transit, personal vehicle, walking, bicycle, and others. The personal vehicle category was used as the reference for the analysis, and results are discussed relative to this category. Interpretation of the results focused on the odds ratios, with p-values used to assess statistical significance. A variable was considered statistically significant at the 95% confidence interval if its p-value was less than or equal to 0.05. The fitted model achieved a prediction accuracy of 91.68%. The personal vehicle was set as the base category for the dependent variables, and results are discussed in relation to this category. Separate multinomial regression models were fitted for each region to assess the results specific to that region.

**Personal Characteristics**
*Age*
The results show a significant correlation between age and mode of transport choice for rural trips. Older adults show a decreased likelihood of using public transit compared to younger individuals aged 11-30, with those aged 31-50 being 0.76 times less likely, 51-70 being 0.49 times less likely, and individuals 71 and above 0.42 times less likely to use public transit than the youngest group. This trend suggests a growing preference for personal vehicles among older rural residents, potentially due to increased car ownership or reduced comfort with public transportation options.

A similar age-related decline is observed in active transportation modes such as walking and biking. Compared to the 11-30 age group, individuals aged 31-50 are 0.57 times less likely to walk and 0.47 times less likely to bike. These findings indicate that as rural residents age, they increasingly favor personal vehicles overactive transportation modes. This shift may be attributed to various factors, including health-related limitations that make walking or biking more challenging for older adults, as well as the convenience and comfort offered by personal vehicles in rural settings.

*Gender*
Gender significantly impacts the mode of transport chosen for rural trips, as evidenced by the p-values being less than 0.05 for several categories (Table 3). Compared to males, females are 0.71 times less likely to use public transit, 0.88 times less likely to walk, 0.30 times less likely to bike, and 0.23 times less likely to use other modes of transportation besides personal vehicles, public transit, walking, and biking.

The lower likelihood of females using active transportation modes such as walking and biking could be due to safety concerns, lack of infrastructure catering to their needs, or sociocultural factors influencing travel behavior. Additionally, the reduced use of other modes by females may stem from limited access to services like ride-sharing or specialized transportation options tailored for women in rural areas. These gender disparities highlight the need for improved safety measures, better infrastructure, and more inclusive transportation services to accommodate the needs of female travelers in rural communities.

**Table 3. Multinomial logistic regression results**

| Variable | Transit | | Walk | | Bicycle | | Other | |
|---|---|---|---|---|---|---|---|---|
| | Odds Ratio | Pvalue | Odds Ratio | Pvalue | Odds Ratio | Pvalue | Odds Ratio | Pvalue |
| *Age* | | | | | | | | |
| 11-30 | | | | | | | | |
| 31-50 | 0.76 | 0.06 | 0.57 | $P < 0.01$ | 0.47 | $P < 0.01$ | 1.99 | $P < 0.01$ |
| 51-70 | 0.49 | $P < 0.01$ | 0.54 | $P < 0.01$ | 0.34 | $P < 0.01$ | 1.94 | $P < 0.01$ |
| 71 and above | 0.42 | $P < 0.01$ | 0.31 | $P < 0.01$ | 0.05 | $P < 0.01$ | 1.15 | 0.29 |
| *Gender* | | | | | | | | |
| Male | | | | | | | | |
| Female | 0.71 | $P < 0.01$ | 0.88 | $P < 0.01$ | 0.3 | $P < 0.01$ | 0.23 | $P < 0.01$ |
| *Education level* | | | | | | | | |
| Highschool graduate or less | | | | | | | | |
| College degree | 0.77 | 0.04 | 1.23 | $P < 0.01$ | 0.75 | 0.06 | 0.78 | $P < 0.01$ |
| Bachelor's degree | 1.13 | 0.32 | 1.87 | $P < 0.01$ | 1.55 | $P < 0.01$ | 0.53 | $P < 0.01$ |
| *Race* | | | | | | | | |
| White | | | | | | | | |
| Black | 4.33 | $P < 0.01$ | 0.92 | 0.35 | 1.53 | 0.11 | 0.68 | 0.03 |
| Hispanic | 1.96 | $P < 0.01$ | 0.76 | 0.01 | 1.63 | 0.05 | 0.84 | 0.33 |
| Other | 1.1 | 0.71 | 1.16 | 0.06 | 1.07 | 0.8 | 1.05 | 0.74 |
| *Household Income* | | | | | | | | |
| less than 25,000 | | | | | | | | |
| 25000-49999 | 0.45 | $P < 0.01$ | 0.97 | 0.57 | 0.7 | 0.04 | 1.3 | 0.01 |
| 50,000-74999 | 0.54 | $P < 0.01$ | 1.12 | 0.04 | 0.52 | $P < 0.01$ | 1.27 | 0.02 |
| 75,000-99999 | 0.35 | $P < 0.01$ | 1.21 | $P < 0.01$ | 0.56 | $P < 0.01$ | 1.31 | 0.02 |
| 100,000 and more | 1.14 | 0.39 | 1.7 | $P < 0.01$ | 0.94 | 0.72 | 1.48 | $P < 0.01$ |
| *Household Size* | 0.87 | $P < 0.01$ | 0.88 | $P < 0.01$ | 0.75 | $P < 0.01$ | 0.92 | $P < 0.01$ |
| *Household Vehicle* | 0.65 | $P < 0.01$ | 0.94 | $P < 0.01$ | 0.92 | 0.06 | 1.14 | $P < 0.01$ |
| *Trip purpose* | | | | | | | | |
| Work | | | | | | | | |
| Non-home based | 0.67 | $P < 0.01$ | 0.43 | $P < 0.01$ | 0.44 | $P < 0.01$ | 1.75 | $P < 0.01$ |
| Shopping | 0.26 | $P < 0.01$ | 0.62 | $P < 0.01$ | 0.61 | 0.05 | 0.74 | $P < 0.01$ |
| Recreation | 0.6 | $P < 0.01$ | 4.69 | $P < 0.01$ | 5.99 | $P < 0.01$ | 2 | $P < 0.01$ |
| Other | 0.57 | $P < 0.01$ | 3.32 | $P < 0.01$ | 2.03 | $P < 0.01$ | 0.89 | 0.27 |
| *Trip Day* | | | | | | | | |
| Weekend | | | | | | | | |
| Weekday | 1.58 | $P < 0.01$ | 1.04 | 0.24 | 1.36 | 0.02 | 1.21 | 0.01 |
| *Trip miles* | 1.00 | $P < 0.01$ | 0.18 | $P < 0.01$ | 0.64 | $P < 0.01$ | 1 | $P < 0.01$ |
| *Census Region* | | | | | | | | |
| Northeast | | | | | | | | |
| Midwest | 0.17 | $P < 0.01$ | 0.64 | $P < 0.01$ | 1.94 | $P < 0.01$ | 1.22 | 0.02 |
| South | 0.23 | $P < 0.01$ | 0.55 | $P < 0.01$ | 0.71 | 0.02 | 0.98 | 0.74 |
| West | 0.38 | $P < 0.01$ | 0.92 | 0.06 | 1.14 | 0.44 | 0.94 | 0.54 |

*Education Level*
Education level plays a significant role in the mode of transport chosen for rural trips, as indicated by the p-values in Table 2. Compared to individuals with a high school education or less, those with a college degree are 0.77 times less likely to use public transit and 0.78 times less likely to use other modes of transport. However, college graduates are 1.23 times more likely to walk.
Individuals with a bachelor's degree or higher education are 1.87 times more likely to walk and 1.55 times more likely to bike compared to those with a high school education or less. Additionally, they are 0.53 times less likely to use other modes of transport.

These findings suggest that higher educational attainment is associated with a greater likelihood of engaging in active transportation modes such as walking and biking. This trend could be attributed to better health awareness, greater environmental consciousness, or access to resources that promote active commuting. Conversely, the reduced reliance on public transit and other modes of transport among higher-educated individuals may be linked to their greater financial capacity to own and maintain personal vehicles.

*Race*
Race significantly influences the mode of transport chosen for rural trips, as shown by the p-values in Table 2. Compared to Whites, Black Americans are 4.33 times more likely to use public transit, 0.68 times less likely to use other modes of transport, and do not show a significant difference in walking. Hispanics are 1.96 times more likely to use public transit, 0.76 times less likely to walk, and 1.63 times more likely to bike compared to Whites.

These results highlight racial disparities in transportation mode choices. The higher likelihood of using public transit among Black Americans and Hispanics may indicate greater reliance on public transportation due to socio-economic factors such as lower vehicle ownership rates and income levels. The lower likelihood of walking among Hispanics could be due to cultural preferences or safety concerns. The increased likelihood of biking among Hispanics suggests that biking might be a more accessible and affordable mode of transportation for this group.

**Household Characteristics**
*Household Size*
The size of a household plays a crucial role in determining the mode of transport for rural trips, as evidenced by the p-values in Table 2. With each additional household member, the likelihood of using public transit decreases by 0.87 times, walking by 0.88 times, biking by 0.75 times, and other modes of transport by 0.92 times.

This trend indicates that larger households are less inclined to use public transit and active transportation modes. The reasons for this could include logistical challenges in coordinating travel for multiple members and a higher reliance on personal vehicles to meet the diverse needs of a larger family. Financial constraints may also limit the ability of larger households to afford multiple transportation options, making personal vehicles a more practical choice for managing daily travel needs.

*Household Vehicles*
The p-values in Table 2 show that the number of vehicles in a household significantly impacts the mode of transport chosen for rural trips. An increase in the number of household vehicles decreases the likelihood of using public transit and walking by 0.65 times and 0.94 times, respectively. Conversely, the presence of more household vehicles increases the likelihood of using other modes of transport by 1.14 times. This suggests that households with more vehicles are more likely to rely on personal and other motorized modes of transport, reducing their dependence on public transit and active transportation. The availability of household vehicles offers greater flexibility and convenience, allowing residents to cover longer distances and access a wider range of destinations more efficiently.

**Trip Characteristics**
*Trip Purpose*
The characteristics of a trip, such as its purpose and timing, significantly influence the choice of transport mode. For work-related trips, public transit is more likely to be used. In contrast, non-home-based, shopping, recreation, and other purpose trips are 0.67, 0.26, 0.60, and 0.57 times less likely to be made using public transit, respectively. Recreational trips, on the other hand, are more likely to involve active transportation modes. These trips are 4.69 times more likely to involve walking and 5.99 times more likely to involve biking, indicating the importance of having safe and accessible infrastructure for walking and biking to support recreational activities.

*Trip Day*
The day of the week also influences the mode of transport chosen. Weekday trips are 1.58 times more likely to involve public transit compared to weekend trips. Additionally, weekday trips are 1.36 times more likely to involve biking and 1.21 times more likely to involve other modes of transport.
This suggests that public transit and active transportation modes are more commonly used for weekday activities, likely due to the regular commuting needs for work and school. Understanding these patterns can help optimize public transit schedules and improve infrastructure to support active transportation during the weekdays.

*Trip Length*
Trip length significantly impacts the mode of transport chosen for rural trips. As the length of the trip increases, the likelihood of using public transit and other modes increases, while the likelihood of using active modes such as walking and biking decreases. Specifically, for each additional mile, a trip is 1.01 times more likely to use public transit and other modes, 0.64 times less likely to involve biking, and 0.18 times less likely to involve walking.
This indicates that longer trips are better suited for public transit or motorized modes of transport, whereas shorter trips are more feasible for active transportation modes. These findings are similar to those of Yang et al. 2018, suggesting that the trip distance and cost increase the chances of using public transit when making a trip. This underscores the need for integrated transportation solutions that offer efficient public transit for longer trips while promoting active transportation for shorter distances.

**Regional Comparison**
A separate multinomial model was created for each of the four census regions. The results from the multinomial logistic regression models for the four census regions—Northeast, Midwest, South, and West—reveal distinct patterns and variations in transportation mode choices across the United States. This section compares these regional differences to comprehensively understand how various factors influence transportation preferences in each region.

Across rural regions in the United States, transportation patterns have several notable similarities. Older adults (71 and above) are consistently less likely to walk, bike, or use public transit, preferring personal vehicles or other modes of transport. Females are generally more inclined to use personal vehicles and are less likely to walk, bike, or use other modes than males. Higher educational attainment is associated with increased use of personal vehicles and decreased use of public transit or other modes, likely due to better job opportunities and higher vehicle ownership among more educated individuals. Black Americans and Hispanics consistently show higher usage of public transit compared to Whites, with Hispanics also having higher bicycle usage in regions like the South and West. Higher household incomes correlate with increased use of personal vehicles and decreased use of public transit, as wealthier households tend to rely more on personal vehicles. The presence of household vehicles

Despite these similarities, rural areas have distinct regional differences influenced by demographic, economic, and infrastructural factors. In the rural Northeast, higher public transit usage is observed among Black Americans and Hispanics, with Black Americans five times more likely and Hispanics nearly four times more likely to use public transit compared to Whites. This higher public transit usage is likely due to the availability of public transit options. Recreational trips in the Midwest also show a higher likelihood of involving walking and biking. This could be due to the availability of recreational areas and a cultural emphasis on outdoor activities. The presence of extensive road networks and less congested areas might also contribute to these patterns.

**CONCLUSIONS**
This study used the 2017 NHTS data to examine the travel behavior of people in rural areas and employed multinomial logistic regression to analyze how the mode choice of transportation is affected by socio-demographics, household, and trip characteristics. Results from the MNL show that older adults (51-70 and 71+ age groups) are 0.49 and 0.42 times less likely to use public transit compared to younger individuals. Females are 0.71 times less likely to use public transit and 0.30 times less likely to bike. College degree holders are 1.23 times more likely to walk, and individuals with a bachelor's degree or higher are 1.87 times more likely to walk and 1.55 times more likely to bike. Black Americans and Hispanics are 4.33 and 1.96 times more likely to use public transit than Whites.

Work-related trips are more likely to involve public transit, while recreational trips are 4.69 times more likely to involve walking and 5.99 times more likely to involve biking. Weekday trips are 1.58 times more likely to use public transit compared to weekend trips. Regional differences show that individuals in the Northeast are significantly more likely to use public transit than those in the Midwest, South, and West. As trip length increases, the likelihood of using public transit rises, while walking and biking become less feasible, suggesting the integration of multimodal options for longer trips.

These findings emphasize the need for targeted strategies, including developing age-friendly transportation options, promoting active commuting through educational programs and infrastructure improvements, and ensuring equitable transit services for minority communities. By considering these factors, transportation systems can become more inclusive, efficient, and responsive to the needs of diverse populations, enhancing mobility for all rural communities.

Future studies could utilize the 2022 NHTS data to determine how travel behaviors in rural areas have changed since the pandemic. Additionally, they could analyze travel patterns by examining how built infrastructure impacts the choice of transportation mode in rural areas—comparing the travel patterns of rural areas close to metropolitan areas with those more isolated ones.

**AUTHOR CONTRIBUTIONS**
The authors confirm their contributions to the paper as follows: study conception and design: Doreen Kobelo, Mohamed Khallafala; data collection: Janerose Matyenyi; analysis and interpretation of results: Janerose Matyenyi, Thobias Sando: draft manuscript preparation: Janerose Matyenyi, Thobias Sando, Doreen Kobelo, Mohamed Khallafala. All authors reviewed the results and approved the final version of the manuscript.

**REFERENCES**

1. Rural Public Transportation Systems | US Department of Transportation. https://www.transportation.gov/mission/health/Rural-Public-Transportation-Systems. Accessed May 18, 2024.

2. Yu, Z., and P. Zhao. The Factors in Residents' Mobility in Rural Towns of China: Car Ownership, Road Infrastructure and Public Transport Services. *Journal of Transport Geography*, Vol. 91, 2021, p. 102950. https://doi.org/10.1016/J.JTRANGEO.2021.102950.

3. Blumenberg, E., J. Paul, and G. Pierce. Travel in the Digital Age: Vehicle Ownership and Technology-Facilitated Accessibility. *Transport Policy*, Vol. 103, 2021, pp. 86–94. https://doi.org/10.1016/J.TRANPOL.2021.01.014.

4. Ratcliffe, M., C. Burd, K. Holder, and A. Fields. Defining Rural at the U.S. Census Bureau. 2016.

5. Census. government (2022). 2020 Census Urban Areas Facts [Online]. United Census Bureau. Available: https://www.census.gov/programs-surveys/geography/guidance/geo-areas/urban-rural/2020-ua-facts.html Accessed May 18, 2024.

6. Geographic Levels. https://www.census.gov/programs-surveys/economic-census/guidance-geographies/levels.html. Accessed Jul. 20, 2024.

7. Pucher, J., and J. L. Renne. *Rural Mobility and Mode Choice: Evidence from the 2001 National Household Travel Survey*.

8. Mattson, J. *RURAL TRANSIT FACT BOOK, 2017*. 2017.

9. Brown, Dennis M. & Stommes, Eileen S., 2004. "Rural Governments Face Public Transportation Challenges and Opportunities," Amber Waves: The Economics of Food, Farming, Natural Resources, and Rural America, United States Department of Agriculture, Economic Research Service, pages 1-2, February.

10. Transportation Institute, M. *Investigating the Determining Factors for Transit Travel Demand by Bus Mode in US Metropolitan Statistical Areas*. 2012.

11. Yanar, T. Understanding the Choice for Sustainable Modes of Transport in Commuting Trips with a Comparative Case Study. *Case Studies on Transport Policy*, Vol. 11, 2023, p. 100964. https://doi.org/10.1016/J.CSTP.2023.100964.

12. Collia, D. V., J. Sharp, and L. Giesbrecht. The 2001 National Household Travel Survey: A Look into the Travel Patterns of Older Americans. *Journal of Safety Research*, Vol. 34, No. 4, 2003, pp. 461–470. https://doi.org/10.1016/J.JSR.2003.10.001.

13. Paul, J. Work from Home Behaviors among U.S. Urban and Rural Residents. *Journal of Rural Studies*, Vol. 96, 2022, pp. 101–111. https://doi.org/10.1016/j.jrurstud.2022.10.017.

14. Tribby, C. P., and D. S. Tharp. Examining Urban and Rural Bicycling in the United States: Early Findings from the 2017 National Household Travel Survey. *Journal of transport and health*, Vol. 13, 2019, pp. 143–149. https://doi.org/10.1016/j.jth.2019.03.015.

15. Mahmoudi, J. Health Impacts of Nonmotorized Travel Behavior and the Built Environment: Evidence from the 2017 National Household Travel Survey. *Journal of Transport & Health*, Vol. 26, 2022, p. 101404. https://doi.org/10.1016/J.JTH.2022.101404.

16. Federal Highway Administration. (2017). 2017 NHTS Data User Guide. (2018.

17. Ao, Y., Y. Zhang, Y. Wang, Y. Chen, and L. Yang. Influences of Rural Built Environment on Travel Mode Choice of Rural Residents: The Case of Rural Sichuan. *Journal of Transport Geography*, Vol. 85, 2020. https://doi.org/10.1016/j.jtrangeo.2020.102708.

18. Yang, L. Modeling the Mobility Choices of Older People in a Transit-Oriented City: Policy Insights. *Habitat International*, Vol. 76, 2018, pp. 10–18. https://doi.org/10.1016/j.habitatint.2018.05.007.

19. Chapter 11 Multinomial Logistic Regression | Companion to BER 642: Advanced Regression Methods. https://bookdown.org/chua/ber642_advanced_regression/multinomial-logistic-regression.html. Accessed Jun. 9, 2024.

20. Bernasco, W., and R. Block. Discrete Choice Modeling.